\documentclass[aps,11pt]{revtex4}
\usepackage{dcolumn}
\usepackage{graphicx}
\usepackage{amsmath}
\usepackage{amsfonts}
\usepackage{amssymb}
\usepackage{psfrag}
\usepackage{wrapfig}
\usepackage{subfigure}
\usepackage{makeidx}
\usepackage{bm}
\usepackage{epsf}
\usepackage{hyperref}
 \usepackage{float}
\usepackage{color}
\usepackage{cases}
\usepackage{booktabs} 
\usepackage{makecell} 
\usepackage{multirow}
\usepackage{bm}

\makeatletter
\newcommand{\eq}[1]{Eq.(\ref{#1})}

\newcommand{\Rmnum}[1]{\uppercase\expandafter{\romannumeral #1}}
\usepackage{amsmath} 
\begin{document} 

\title{Perturbations of massless external fields on magnetically charged black holes in string-inspired Euler-Heisenberg theory}

\author{Xufen Zhang$^{1}$\footnote{xfzhangyzu@126.com}, De-Cheng Zou$^{2}$\footnote{Corresponding author: dczou@jxnu.edu.cn}, Chao-Ming Zhang$^{1}$\footnote{zcm843395448@163.com}, Ming Zhang$^{3,4}$\footnote{mingzhang0807@126.com} \\ and Rui-Hong Yue$^{1}$\footnote{Corresponding author: rhyue@yzu.edu.cn}}

\address{
$^{1}$Center for Gravitation and Cosmology, College of Physical Science and Technology, Yangzhou University, Yangzhou 225009, China \\
$^{2}$College of Physics and Communication Electronics, Jiangxi Normal University, Nanchang 330022, China \\
$^{3}$Faculty of Science, Xihang University, Xi'an 710077 China \\
$^{4}$National Joint Engineering Research Center for Special Pump System Technology, Xihang University, Xi’an 710077, China}

\date{\today}

\begin{abstract}
\indent

In this paper, we study the perturbations of massless scalar and electromagnetic fields on the magnetically charged black holes in string-inspired Euler-Heisenberg theory. We calculate the quasinormal frequencies (QNFs) and discuss influences of black hole magnetic charge $Q_m$, coupling parameter $\epsilon$ and angular momentum $l$ on QNFs, emphasizing the relationship between these parameters and QNMs behavior. We find these results obtained through the AIM method are in good agreement with those of obtained by WKB method. The greybody factor is calculated by WKB method. The effects of these parameters $Q_m$ and $\epsilon$ on the greybody factor are also studied. 
\end{abstract}


\maketitle

\section{Introduction}
\label{intro}

As a nonlinear extension of quantum electrodynamics (QED), the Euler-Heisenberg (EH) Lagrangian formulated in 1936 \cite{Heisenberg:1936nmg} provides a classical approximation superior to Maxwell theory under strong-field conditions where vacuum polarization becomes significant. This framework models the vacuum as a dynamically polarizable medium, with polarization/magnetization arising from virtual charge clouds around real charges/currents \cite{Obukhov:2002xa}. This theory not only provides a more precise classical approximation of QED than Maxwell electrodynamics under strong-field conditions, but also serves as a foundational tool for studying nonlinear phenomena in both astrophysics and cosmology.
Building on its unique physical properties, the first EH black hole solution-an anisotropic Reissner-Nordström-like magnetically charged configuration with dyon degrees of freedom was derived in 1956 \cite{Yajima:2000kw}. Subsequent studies focus on electrically charged solutions \cite{Yajima:2000kw,Ruffini:2013hia}, rotating solutions \cite{Breton:2019arv,Amaro:2022yew}, and frameworks based in modified gravity theories \cite{Guerrero:2020uhn,Nashed:2021ctg}. Very recently, inspired by string theory and Lovelock theory, Ref. \cite{Bakopoulos:2024hah} proposed an extension of Einstein-Maxwell-dilaton theory by coupling dilaton with EH electrodynamics in a specified way.
Then, some authors investigated the effects of particle motion and gravitational lensing phenomenon on the magnetically charged black holes \cite{Yasir:2025npe,Vachher:2024fxs}, and black hole shadow \cite{Xu:2024gjs}. Subsequently, Jiang et al. \cite{Jiang:2024njc} described geometrically thin and optically thick accretion disks around the magnetically charged black holes.

Quasinormal modes (QNMs) are fundamental dissipation signatures of spacetime perturbations, particularly significant in black hole systems where energy loss to infinity or absorption by the event horizon dominates. The analysis of QNMs provides valuable insights into black hole spacetime geometry and could be instrumental in gravitational wave observations, as well as in exploring fundamental symmetries in gauge/gravity correspondence. Note that Cho et al. \cite{Cho:2009cj,Cho:2011sf} presented a new method known as the asymptotic iteration method (AIM), which 
employs a systematic iterative approach to solve the derivation of accurate approximations to the quasinormal frequencies(QNFs). We are going to employ the AIM method to solve the perturbation equation numerically.
In addition, we shall utilize the WKB method \cite{Iyer:1986np,Iyer:1986nq, Konoplya:2003ii}, a widely recognized and established technique, to estimate the QNFs of the black holes. By comparing the results obtained via the two approaches, we aim to validate and strengthen our findings. 
Another important concept for a black hole perturbation system is the greybody factor \cite{Konoplya:2019ppy,Konoplya:2011qq,Konoplya:2019hlu,Gogoi:2023fow}. It describes the modification of the spectrum
of radiation as it escapes the black hole’s gravitational well.
In the context of gravitational wave astronomy, the interplay between QNFs and the greybody factor establishes a self-consistent framework for interpreting compact object mergers. QNFs govern the post-merger ringdown phase via their resonant frequencies and damping rates, whereas the greybody factor characterizes the anisotropic emission of gravitational waves during the complete inspiral-merger-ringdown sequence \cite{Oshita:2023cjz}.

In light of these, we plan to study the QNFs and greybody factor of test external fields on the magnetically charged black hole in string-inspired Euler–Heisenberg theory. This paper is constructed as follows. In Section \ref{sec2}, we review the black hole solution briefly. Then we derive the master equations for test massless scalar and electromagnetic fields in Section \ref{sec3}. In Section \ref{sec4}, we solve the QNFs with AIM and WKB methods, and study the effects of the black hole parameters on the QNFs. In Section \ref{sec5}, we calculate the greybody factor with the WKB method. The conclusions and and discussion are given in Section \ref{sec6}.

\section{Background solution}
\label{sec2}

Inspired by string theory and Lovelock theory, Bakopoulos et al.\cite{Bakopoulos:2024hah}  recently propose the Einstein-Maxwell-dilaton theory including a dilaton-coupled
nonlinear Euler-Heisenberg term
\begin{eqnarray}
S=\frac{1}{16\pi}\int  d^4x\sqrt{-g}\Big[R-2\nabla^\mu\phi\nabla_\mu\phi-e^{-2\phi}F^2-f(\phi)\left(2\alpha F^\mu_{~\nu} F^\nu_{~\rho} F^\rho_{~\delta} F^\delta_{~\mu}-\beta F^4\right)\Big],\label{action}
\end{eqnarray}
where $R$ denotes the scalar curvature, $f(\phi)$ is a coupling function depending on scalar field $\phi$ and usual Faraday scalar 
 $F^2=F_{\mu\nu}F^{\mu\nu}$ and $F^4=F_{\mu\nu}F^{\mu\nu}F_{\rho\delta}F^{\rho\delta}$, where $F_{\mu\nu}$ stands for the usual field strength $F_{\mu\nu}=\partial_\mu A_{\nu}-\partial_{\nu}A_{\mu}$. Here $\alpha$, $\beta$ are coupling constants of the theory.  
If one sets $\alpha=\beta=0$, the model will reduce to the standard Einstein-Maxwell-dilaton theory. 

Assuming the function $f(\phi)$ taking the following form
\begin{eqnarray}
 f(\phi)=-[3\text{cosh}(2\phi)+2]\equiv-\frac{1}{2}(3e^{-2\phi}+3e^{2\phi}+4),\label{phi}
\end{eqnarray}
the action \eqref{action} admits magnetically charged black
hole with the metric \cite{Bakopoulos:2024hah}
\begin{eqnarray}
 ds^2&=&-h(r) \,dt^2 + \frac{1}{h(r)} \, dr^2 + R(r)^2 \,( d\theta^2 + \sin^2 \theta \, d\varphi^2),\label{oldmetric}\\
h(r)&=&1-\frac{2 M}{r}-\frac{2(\alpha-\beta)Q_m^4}{r^3(r-Q_m^2/M)^3}, \quad
R(r)=r\left(r-\frac{Q_m^2}{M}\right),\nonumber\\
\phi (r)&=&-\frac{1}{2}\ln \left(1-\frac{Q_m^2}{M r}\right), \quad A_\mu=(0, 0, 0, Q_m\cos\theta),\label{Qm}\label{fRsolution}
\end{eqnarray}
where $M$ and $Q_m$ are the mass and magnetic charge of this black hole, respectively. We define $\epsilon=\alpha-\beta$ for simplify.

It is noteworthy that,  in the limit $\epsilon=0$,  the solution \eqref{fRsolution} reduces to the Gibbons-Maeda-Garfinkle-Horowitz-Strominger (GMGHS or GHS) black holes \cite{Gibbons:1987ps,Garfinkle:1990qj} and further becomes the Schwarzschild solution when $Q_m=0$. Until now, these works have generated enormous interest in the GMGHS or GHS black holes \cite{Ferrari:2000ep,Chen:2004zr,Shu:2004fj,Karimov:2018whx}. In addition,  Ref.\cite{Bakopoulos:2024hah} shows that the solution \eqref{fRsolution} with $\epsilon\neq0$ describes a black hole with a single horizon when $\epsilon=1$, while for $\epsilon=-1$, the black hole horizons can range from two to none. These imply the magnetically charged black hole possesses different horizon structures. Therefore, we consider these two branches solutions ($\epsilon>0$ and $\epsilon<0$), respectively.

From \eq{oldmetric} and \eq{fRsolution}, we can also rewrite the old metric \eqref{oldmetric} to a new spherically symmetric metric ansatz
\begin{eqnarray}
  ds^2 = -A(r) \,dt^2 + \frac{1}{B(r)} \, dr^2 + r^2 (d\theta^2 + \sin^2 \theta \, d\varphi^2),\label{metric}
\end{eqnarray}
and the magnetically charged black hole solution are obtained as
\begin{eqnarray}
A(r)&=&1-\frac{4 M^2}{Q_m^2+\sqrt{Q_m^4+4 M^2 r^2}}-\frac{2\epsilon Q_m^4}{r^6},\nonumber\\
B(r)&=&1
- \frac{Q_{m}^{4} + 4M^{2}r^{2}}{r^2(Q_{m}^{2} + \sqrt{Q_{m}^{4} + 4M^{2}r^{2}})} +\frac{Q_m^4}{4M^2r^2}
- \frac{\epsilon Q_{m}^{4}(Q_{m}^{4} + 4M^{2}r^{2})}{2M^2r^{8}},\nonumber\\
\phi (r)&=&-\frac{1}{2}\ln \left(\frac{\sqrt{Q_m^4+4 M^2 r^2}-Q_m^2}{\sqrt{Q_m^4+4 M^2 r^2}+Q_m^2}\right)\label{solution}.
\end{eqnarray}

\section{Wave equations and perturbations}
\label{sec3}

In this section, we will show the master equations of various external fields, including the scalar field and electromagnetic field around magnetically charged black hole. 

\subsection{Scalar field perturbation}

This scalar perturbation can be treated as a probe into the background with a fixed geometry and perturbation equation can be written as 
\begin{eqnarray}   
\Box\Phi=\frac{1}{\sqrt{-g}}\partial_{\mu}\left(\sqrt{-g}g^{\mu\nu}\partial_{\nu}\Phi\right)= 0\label{eqscalar}.
 \end{eqnarray}  
Considering the line element \eq{metric} in \eq{eqscalar}, we can apply the separation of variables as follows  
\begin{eqnarray}  
\Phi(t, r, \theta, \varphi) = e^{-i \omega t} \frac{\psi_s(r)}{r}Y(\theta, \varphi),
 \end{eqnarray}   
and then the perturbed field equation for the radial
part in the tortoise coordinate can be written as
\begin{eqnarray} 
\frac{d^2 \psi_s(r_*)}{dr_*^{2}} + \left[ \omega^2 - V_s(r) \right] \psi_s(r_*) = 0\label{E},  
\end{eqnarray} 
where the tortoise coordinate $r_*$ is defined as follows:
\begin{eqnarray} 
dr_*=\frac{1}{\sqrt{A(r)B(r)}}dr. \label{rstar}
\end{eqnarray}
Here $\psi_s(r)$ is radial wave function, $\omega$ is the frequency of $\Phi$, $l$ corresponds to the multipole moment of the black hole’s QNMs and $V_s(r)$ is the effective potential 
\begin{eqnarray}  
V_s(r)=\frac{ {A}'(r)B(r)+A(r){B}'(r)}{2r}+\frac{A(r)}{r^2}l(l+1). \label{potVs} 
\end{eqnarray}
Evidently, this effective potential $V_s(r)$ depends on the black hole background as well as multipole moment $l$.

\begin{figure}[htb]
\centering
\subfigure[$\epsilon=1$ and $l=0$ ]
{\label{fig11} 
\includegraphics[width=2in]{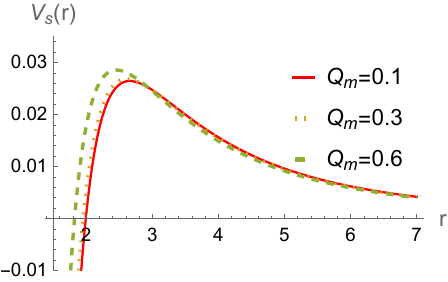}}
\hfill
\subfigure[$\epsilon=1$ and $Q_m=0.3$]
{\label{fig12} 
\includegraphics[width=2in]{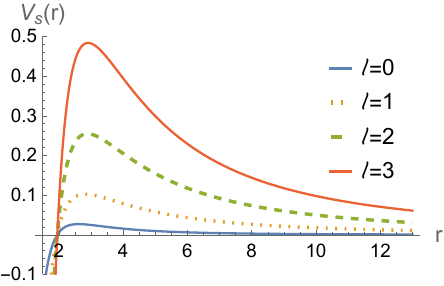}}
\hfill
\subfigure[$Q_m=0.3$ and $l=0$ ]
{\label{fig13}
\includegraphics[width=2in]{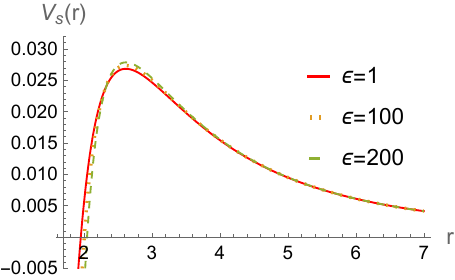}}
\caption{\textbf{The effective potential $V_s(r)$ for massless scalar field perturbation on black hole with $M=1$ and $\epsilon>0$}.}\label{fig1}
\end{figure}

\begin{figure}[htb]
\centering
\subfigure[$\epsilon=-1$ and $l=0$]
{\label{fig21} 
\includegraphics[width=2in]{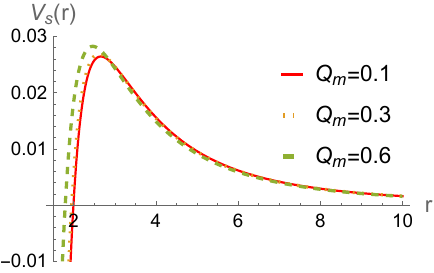}}
\hfill
\subfigure[$\epsilon=-1$ and $Q_m=0.3$]
{\label{fig22} 
\includegraphics[width=2in]{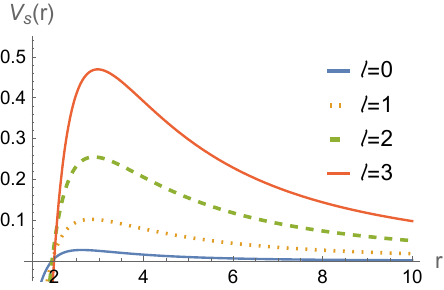}}
\hfill
\subfigure[$Q_m=0.3$ and $l=0$]
{\label{fig23} 
\includegraphics[width=2in]{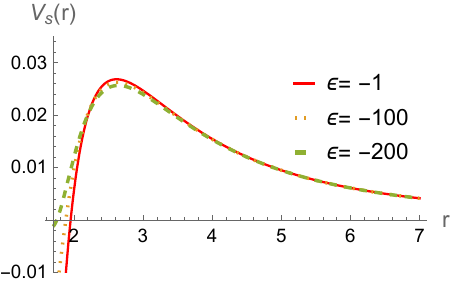}}
\hfill
\caption{\textbf{The effective potential $V_s(r)$ for massless scalar field perturbation on black hole with $M=1$ and  $\epsilon<0$}.}\label{fig2}
\end{figure}

\subsection{Electromagnetic field perturbation}

The propagation of massless electromagnetic field in a curved spacetime, minimally coupled to the geometry, is driven by Maxwell’s equations
\begin{eqnarray}
  \frac{1}{\sqrt{-g}} \, \partial_\mu \left( \sqrt{-g} \, g^{\sigma\mu}g^{\rho\nu} F_{\rho\sigma}\right)=0, \label{eqmax}
\end{eqnarray}
where $F_{\rho\sigma}=\partial_{\rho}A_{\sigma}-\partial_{\sigma}A_{\rho}$  is the field strength tensor, and $A_{\rho}$ is the vector potential  of the perturbed electromagnetic field which can be decomposed as
\begin{eqnarray}
  A_\rho(t, r, \theta, \varphi) = \sum_{l,m} e^{-i\omega t}
\begin{bmatrix}
0 \\
0 \\
h_0(r)\frac{1} {\sin \theta} \frac{\partial Y_{l,m}}{\partial \varphi} \\
-h_0(r) \sin \theta \frac{\partial Y_{l,m}}{\partial \theta}
\end{bmatrix}
+ \sum_{l,m} e^{-i\omega t}
\begin{bmatrix}
h_1(r) Y_{l,m} \\
h_2(r) Y_{l,m} \\
h_3(r) \frac{\partial Y_{l,m}}{\partial \theta} \\
h_3(r) \frac{\partial Y_{l,m}}{\partial\varphi}
\end{bmatrix}. \label{eqA}
\end{eqnarray}
Here, $Y_{l,m}(\theta, \varphi) $ are spherical harmonics, and $l$ and $m$ are the angular and the azimuthal quantum numbers respectively. The first column in \eq{eqA} is the axial component with parity $(-1)^{l+1} $ and the second term is the polar mode with parity $(-1)^l$.

Substituting the axial and the polar modes of the electromagnetic perturbations into \eq{eqmax}, we can obtain 
\begin{eqnarray}
\frac{d^2\Psi_e(r_*)}{dr_*^2} + \left(\omega^2 - V_{e}(r)\right)\Psi_e(r_*) = 0,\label{e}
\end{eqnarray}
where the potential is
\begin{eqnarray}
V_{e}(r) = A(r) \frac{l(l+1)}{r^2} \label{potVe} ,
\end{eqnarray}
and the function \( \Psi_e(r) \) takes different forms for the two modes
\begin{eqnarray}
&&\Psi_e(r) = h_0(r), \quad \quad \quad  \textit{odd parity} ,\\
&&\Psi_e(r) = -\sqrt{\frac{B(r)}{A(r)}}\frac{r^2}{l(l+1)} \left( i\omega h_2(r) + \frac{dh_1(r)}{dr} \right) ,\quad \quad \textit{even parity} .
\end{eqnarray}

\begin{figure}[H]
\centering
\subfigure[$\epsilon=1$ and $l=1$]
{\label{fig31} 
\includegraphics[width=2in]{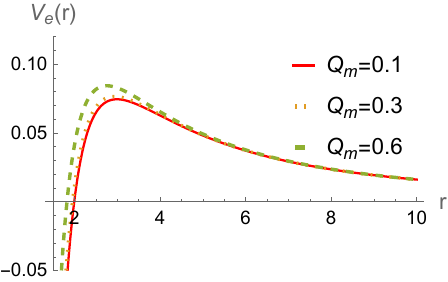}}
\hfill
\subfigure[$\epsilon=1$ and $Q_m=0.3$]
{\label{fig32} 
\includegraphics[width=2in]{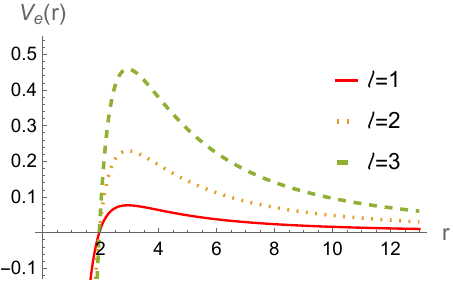}}
\hfill
\subfigure[$l=1$ and $Q_m=0.3$]
{\label{fig33}
\includegraphics[width=2in]{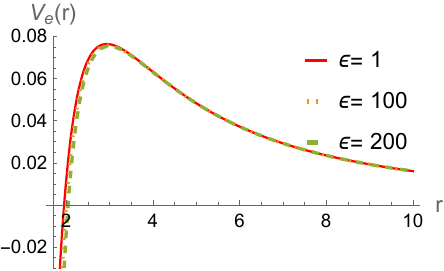}}
\caption{\textbf{The effective potential $V_e(r)$ for massless electromagnetic field on black hole with $M=1$ and $\epsilon>0$}.}\label{fig3}
\end{figure}

\begin{figure}[H]
\centering
\subfigure[$\epsilon=-1$ and $l=1$]
{\label{fig41} 
\includegraphics[width=2in]{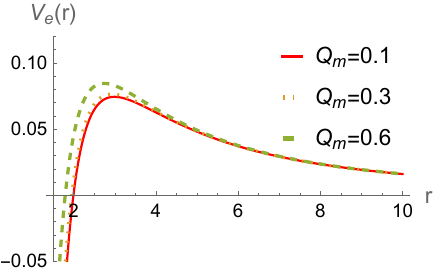}}
\hfill
\subfigure[$\epsilon=-1$ and $Q_m=0.3$]
{\label{fig42} 
\includegraphics[width=2in]{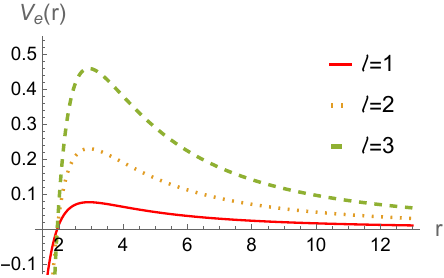}}
\hfill
\subfigure[$l=1$ and $Q_m=0.3$]
{\label{fig43} 
\includegraphics[width=2in]{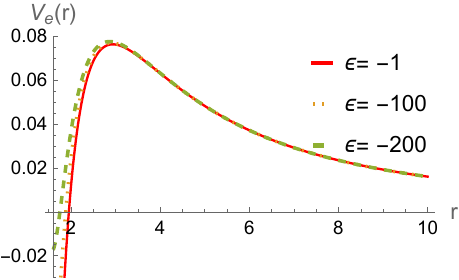}}
\hfill
\caption{\textbf{The effective potential $V_e(r)$ for massless electromagnetic field on black hole with $M=1$ and  $\epsilon<0$}.}\label{fig4}
\end{figure}

Note that the solutions \eqref{solution} are invariant under the following rescaling: $r/M\rightarrow r$, $Q_m /M\rightarrow Q_m$, and  $\epsilon/ M^2\rightarrow \epsilon$. 
For a better analysis of the potential function's behavior, 
We will set $M=1$ throughout the paper and leave $\epsilon$ and $Q_m$ free without loss of generality. The effective potentials $V_s(r)$ and $V_e(r)$ are plotted in Figs.\ref{fig1}-\ref{fig4}. 
Figs.\ref{fig1} and \ref{fig2}, we can see that the height of the potentials $V_s(r)$ increase as charge $Q_m$ and multipole moment $l$ increase in both cases ($\epsilon<0$ and $\epsilon>0$). However, the variation of \textbf{different} $\epsilon$ for these effective potentials is very small. The similar phenomena also appear for the effective potentials $V_e(r)$ of massless electromagnetic field. In addition, the effective potentials are always positive in Figs.\ref{fig1}-\ref{fig4}, indicating that the system is stable under the perturbations of the scalar and electromagnetic fields.

\section{Quasinormal Mode frequencies}
\label{sec4}

Based on the effective potentials obtained above, we can study the QNFs of magnetically charged black holes in string-inspired Euler-Heisenberg theory. QNFs $\omega$ are proper oscillation frequencies corresponding to the solutions of perturbed field equations when appropriate boundary conditions are applied before performing the calculations
\begin{eqnarray}
&&\psi(r_{*}) \sim e^{-i \omega r_{*}}, \quad r_{*} = -\infty,\\
&&\psi(r_{*}) \sim e^{i \omega r_{*}}, \quad r_{*} = +\infty, \label{eq:Phi}
\end{eqnarray}
where purely ingoing modes occur at $r_{*} \to -\infty$ (at the event horizon) and purely outgoing modes occur at $r_{*} \to +\infty$ (at the spatial
infinity). 

In general, the perturbed field equations \eqref{E} and \eqref{e} cannot be derived analytically. In this section, we introduce two different methods, \textit{Asymptotic Iteration Method(AIM)} and \textit{6th-order WKB approximation method(WKB)} to calculate QNFs of magnetically charged black holes and cross-check these results. To streamline the discussion, we provide a concise overview of the WKB and AIM methods in Appendices A and B, as these techniques are well-established in the research. We also calculate the percentage deviation $\Delta_{AW}$ of QNFs obtained via the AIM and WKB methods. The relative error $\Delta_{AW}$ between two methods is defined by
\begin{eqnarray}
\Delta_{AW}=\frac{|\omega_{AIM}-\omega_{WKB}|}{|\omega_{WKB}|}\times 100\%.
\end{eqnarray}

Taking different values of $Q_m$ for two branch solutions, these fundamental QNFs $(n=0)$ are shown in Tables \ref{Table1} and \ref{Table2} for various massless field perturbations.  One can see that these fundamental QNFs obtained by numerical methods agree well with each other for each branch of these black holes. From Table \ref{Table1}, 
the relative error $\Delta_{AW}$ decreases significantly as $l$ increases. These results are consistent with the characteristic features of the WKB method.

\begin{figure}[H]
\centering
\subfigure[$\epsilon=1$]
{
\includegraphics[width=2.3in]{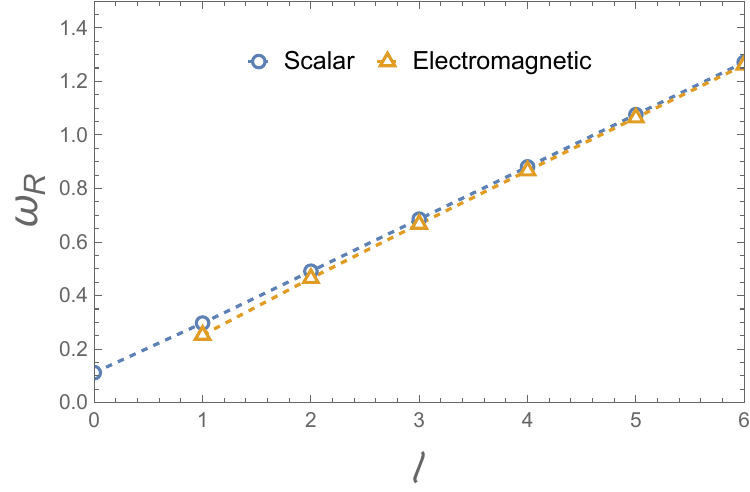}}
\subfigure[$\epsilon=1$]
{ 
\includegraphics[width=2.4in]{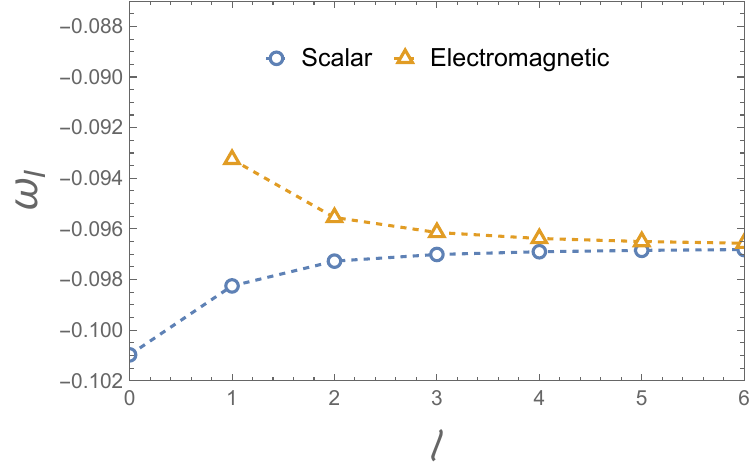}}
\subfigure[$\epsilon=-1$]
{ 
\includegraphics[width=2.3in]{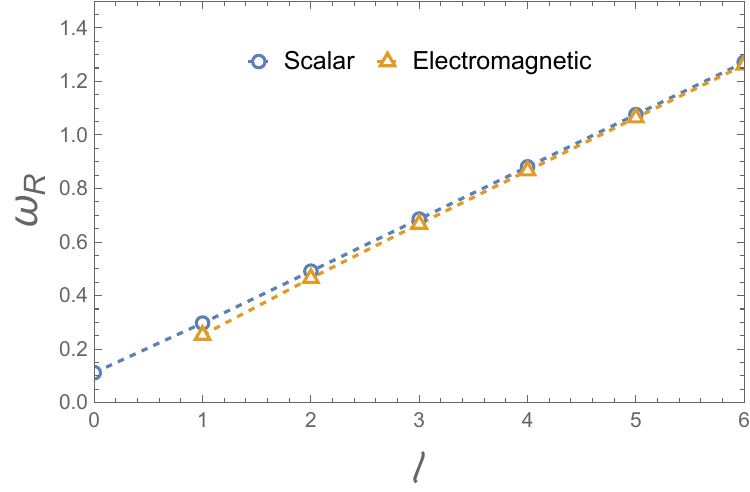}}
\subfigure[$\epsilon=-1$]
{ 
\includegraphics[width=2.4in]{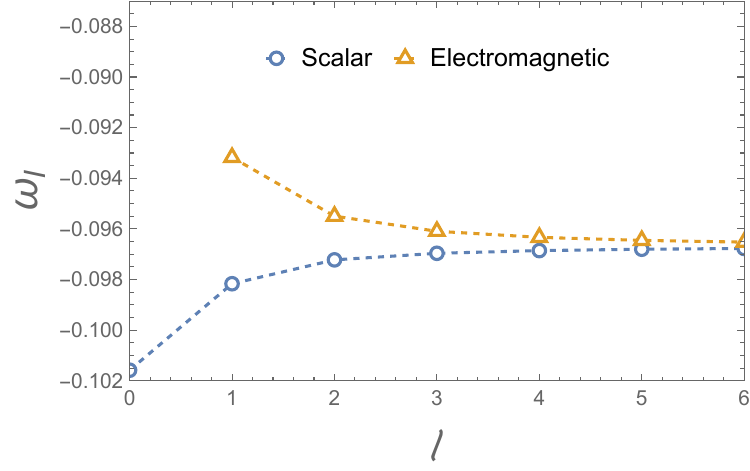}}
\caption{Variation of fundamental QNFs with respect to the \textbf{different} $l$ with $M=1$ and $Q_m=0.3$.}\label{figl1}
\end{figure}

When further increasing $l$, the gap among the  fundamental QNFs for massless perturbing fields gradually tends to be smaller, see Fig.\ref{figl1}. This is because for large, the dominant terms in all the effective potentials \eqref{potVs} and \eqref{potVe} are the terms $l^2$, which have the same formula in all cases.

\begin{table*}[h!]
\caption{The fundamental QNFs $(n=0)$ for scalar perturbation with $M=1$.}\label{Table1}
\resizebox{\linewidth}{!}{
\begin{tabular}{|m{1em}|m{2em}|c|c|c|c|c|c|} \hline
 & & \multicolumn{3}{|c|}{$\epsilon=1$}  &  \multicolumn{3}{|c|}{$\epsilon=-1$} \\ \hline
$l$ &  	$Q_m$ & AIM  &  WKB  & $\Delta_{AW}$   &  AIM  &  WKB   & $\Delta_{AW}$ \\ \hline
 \multirow{3}{*}{0}& 0.1 &   $0.110562-0.105038i$  &  $0.110661 - 0.100865 i$   &2.78754\%   & $0.110562 -0.105037i$        & $0.110663 - 0.100869 i$  & 2.78412 \%  \\ 
& 0.3 &  $0.112132-0.105410i$  & $0.112306 - 0.100973 i$    & 2.94067\%   & $0.112159-0.105337i$   &  $0.112202 - 0.101578 i$  &   $2.48424\%$   \\  
 & 0.5 &  $0.115402-0.106452i$ &$0.116533 - 0.0984805 i$   &  $5.27728\%$ &$0.115647-0.105672i$  & $0.115833 - 0.103841  i$ & $1.18288\%$ \\ 
 &  0.6 &  $0.117741-0.107542i$ &$0.116024 - 0.117945  i$   &  $6.37286\%$ &$0.118292-0.105480i$  & $0.11782 - 0.105505 i$ & $0.29896\%$ \\ \hline
 \multirow{3}{*}{1}&  0.1  &   $0.293432 -0.097712i$  &  $0.293406 - 0.097812  i$   &0.0335086\%   & $0.293432 -0.0977110i$        & $0.293406 - 0.0978113  i$  & 0.0334659\%  \\ 
& 0.3 &  $0.297489-0.098155i$  & $0.29746 - 0.0982498  i$    &0.0316443\%   & $0.297510 -0.098099i$   &  $0.297498 - 0.0981848  i$  &   $0.0276553\%$   \\ 
&  0.5 &  $0.306120 -0.099281i$ &$0.306016 - 0.0994115  i$   &  $0.0519862\%$ &$0.306319-0.0986953i$  & $0.306398 - 0.0986996i$ & $0.0377309\%$ \\ 
&  0.6 &  $0.312483 -0.100351i$ &$0.312211 - 0.100587  i$   &  $0.109943\%$ &$0.312951-0.0986953i$  & $0.313167 - 0.0988303 i$ & $0.0783964\%$ \\ \hline
\end{tabular}}
\end{table*}

\begin{table*}[h!]
\caption{The fundamental QNFs $(n=0)$ for electromagnetic perturbation with $M=1$.}\label{Table2}
\resizebox{\linewidth}{!}{
\begin{tabular}{|m{1em}|m{2em}|c|c|c|c|c|c|} \hline
 & & \multicolumn{3}{|c|}{$\epsilon=1$}  &  \multicolumn{3}{|c|}{$\epsilon=-1$} \\ \hline
$l$ &  	$Q_m$ & AIM  &  WKB  & $\Delta_{AW}$   &  AIM  &  WKB   & $\Delta_{AW}$ \\ \hline
 \multirow{3}{*}{1}& 0.1 &  $0.248737 -0.0925503i$  &  $0.248665 - 0.0927009   i$     & 0.0628748\%   &  $0.248737 -0.0925489i$  &  $0.248666 - 0.0927  i$     & 0.0630446\%   \\ 
& 0.3 &  $0.252604 -0.0930818i $&$0.252514 - 0.0932532 i$    & 0.121663\%   & $0.252646 -0.0930246i$   &  $0.252592 - 0.0931737 i$  &   $0.0588877\%$   \\  
 & 0.5 &  $0.260752 -0.0944050i$ &$0.260503 - 0.0946895i$   &  $0.227963\%$ &$0.261178 -0.0937916i$  & $0.261284 - 0.0938366 i$ & $0.0417106\%$  \\ 
 &  0.6 &  $0.306120 -0.0992805i$ &$0.306016 - 0.0994115  i$   &  $0.0519862\%$ &$0.306319-0.0986953i$  & $0.306398 - 0.0986996i$ & $0.0377309\%$ \\ \hline
 \multirow{3}{*}{2}&  0.1  &   $0.458394-0.0950606i$  &  $0.458392 - 0.0950673  i$   &0.00151693\%   & $0.458394 -0.0950600i$        & $0.458392 - 0.0950673   i$  & 0.00165983\%  \\ 
& 0.3 &  $0.464925-0.0955397i$  & $0.464922 - 0.0955477    i$    &0.00178846\%   & $0.464967 -0.0954883i$  & $0.464922 - 0.0955477   i$    &0.0156059\%     \\ 
&  0.5 &  $0.478801, -0.0967266i$ &$0.47879 - 0.0967427   i$   &  $0.00394474\%$ &$0.479212 -0.0961876i$  & $0.479218 - 0.0961862 i$ & $0.00130171\%$ \\ 
&  0.6 &  $0.488995-0.0978241i$ &$0.488972 - 0.0978537   i$   &  $0.00759593\%$ &$0.490021 -0.0964230i$  & $0.490039 - 0.0964063 i$ & $0.00494179\%$ \\ \hline
\end{tabular}}
\end{table*}

\subsection{\textbf{\texorpdfstring{$Q_m$}--dependence} }

\begin{figure}[H]
\centering
\subfigure[$\epsilon=1$]
{ \label{figQ11}
\includegraphics[width=2.3in]{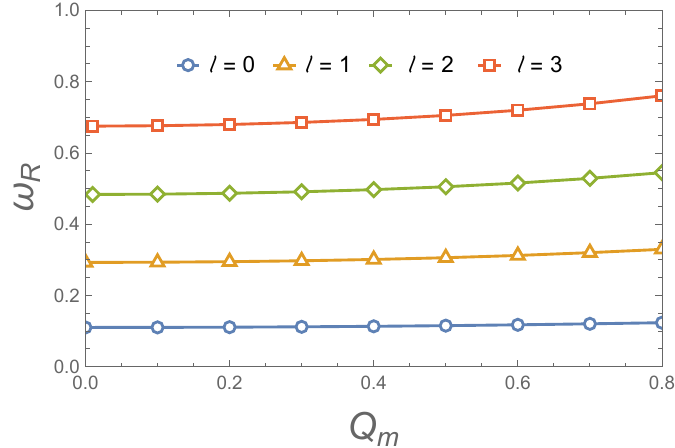}}
\subfigure[$\epsilon=1$]
{\label{figQ12} 
\includegraphics[width=2.35in]{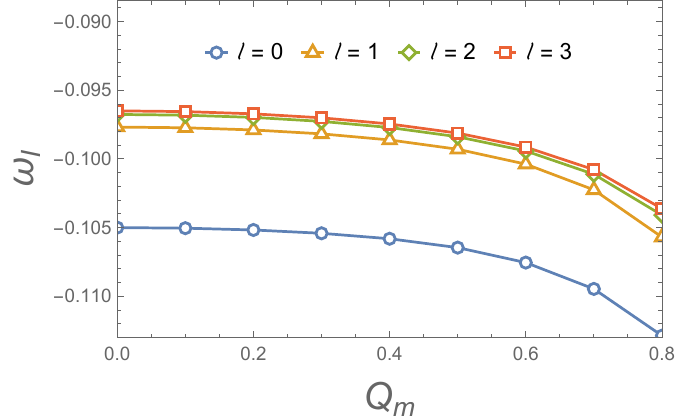}}
\subfigure[$\epsilon=-1$]
{ \label{figQ13}
\includegraphics[width=2.3in]{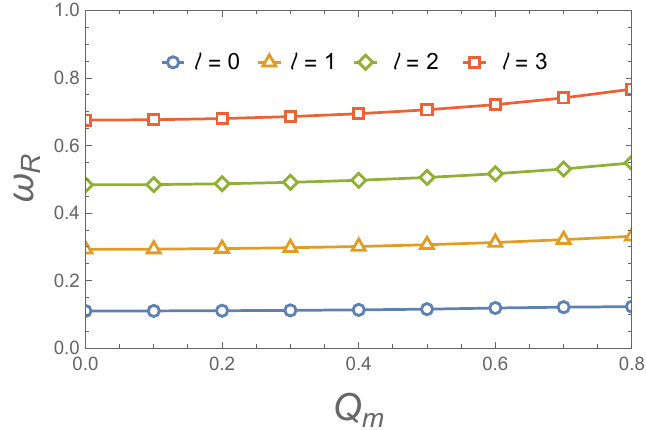}}
\subfigure[$\epsilon=-1$]
{\label{figQ14} 
\includegraphics[width=2.35in]{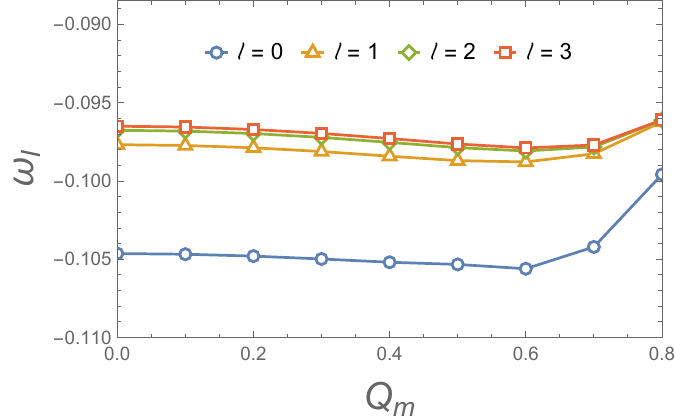}}
\caption{Variation of scalar fundamental QNFs  with respect to the magnetic charge $Q_m$ with $M=1$.}\label{figQ1}
\end{figure}

\begin{figure}[H]
\centering
\subfigure[$\epsilon=1$]
{ \label{figQ21}
\includegraphics[width=2.3in]{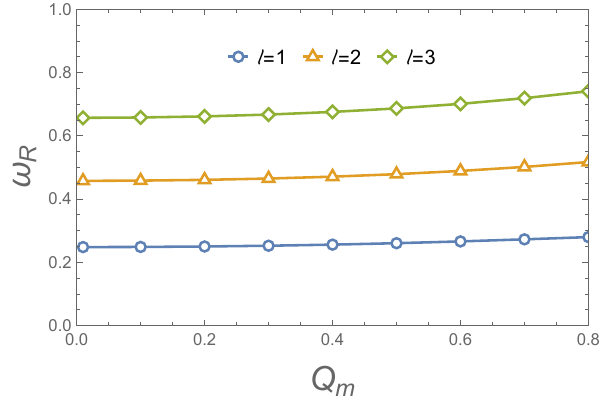}}
\subfigure[$\epsilon=1$]
{ \label{figQ22}
\includegraphics[width=2.4in]{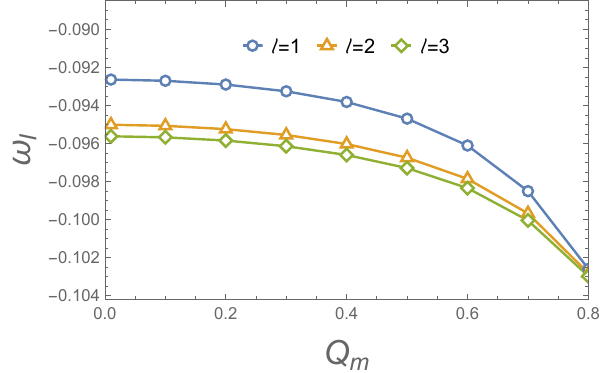}}
\subfigure[$\epsilon=-1$]
{
\includegraphics[width=2.3in]{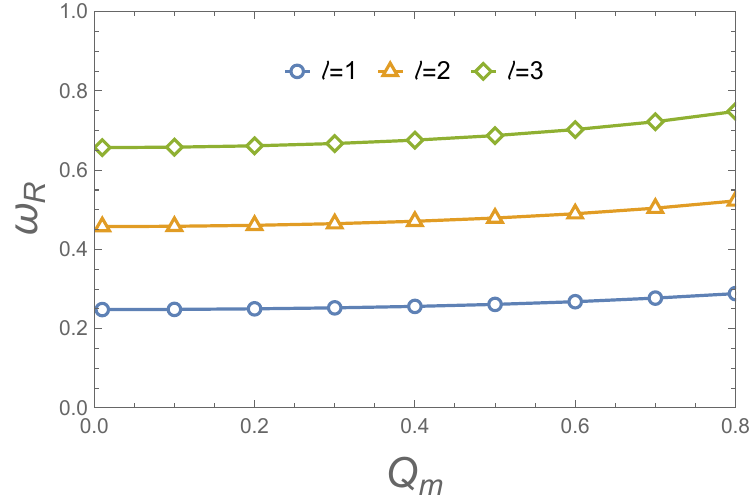}}
\subfigure[$\epsilon=-1$]
{\label{figQ24} 
\includegraphics[width=2.4in]{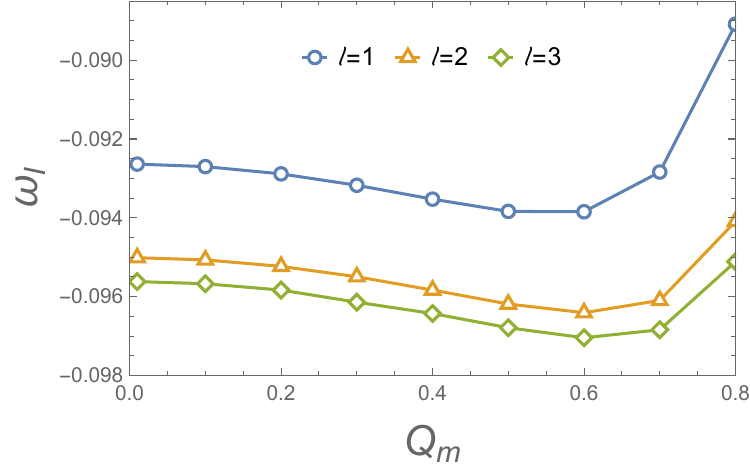}}
\caption{Variation of electromagnetic fundamental QNFs with respect to the \textbf{different} $Q_m$ with  $M=1$.}\label{figQ2}
\end{figure}

Now we analyze the influence of the magnetic charge $Q_m$ on the QNFs in various external field perturbations. These fundamental QNFs with respect to different values $Q_m$ are plotted in Figs.\ref{figQ1} and \ref{figQ2}. For black holes from different branches, these QNFs exhibit distinct characteristics under each type of test field perturbation with increase of $Q_m$. However, for black holes within the same branch, these QNFs for scalar and electromagnetic perturbations show the same trend of variation as  $Q_m$ change. 

Taking scalar field perturbation for example, the real QNFs for $\epsilon=1$ increase as $Q_m$ increases, and the damping rate or decay rate of perturbed field increases significantly with an increase of $Q_m$, see Figs. \ref{figQ11} and \ref{figQ12}. With regard to $\epsilon=-1$ in Figs. \ref{figQ13} and \ref{figQ14}), the real QNFs also increase as $Q_m$ increases. However, the damping rate increases with the growth of $Q_m$, and then reach a maximum decay rate occurring at $Q_m=0.6$. As $Q_m$ surpasses this point, the damping rate begins to decrease. 
The similar behaviors also appear for the electromagnetic field perturbation, see Fig.\ref{figQ2}.

\begin{figure}[H]
\centering
\subfigure[Perturbed scalar field]{\label{figQ31} 
\includegraphics[width=2.3in]{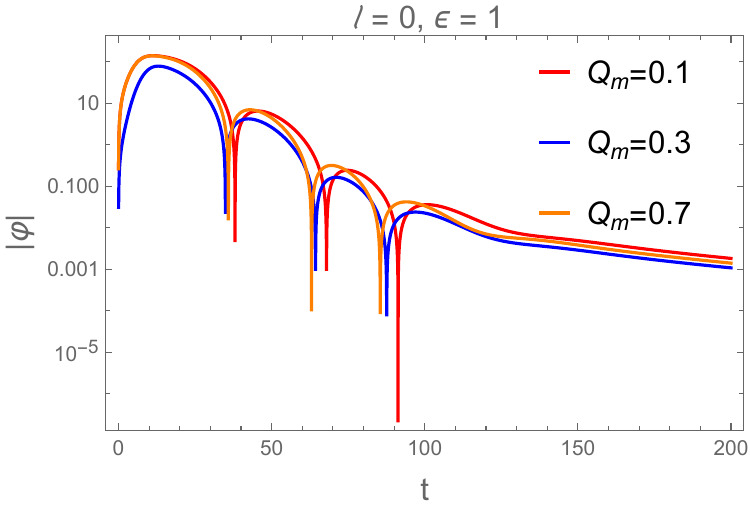}}
\subfigure[Perturbed scalar field]{\label{figQ32} 
\includegraphics[width=2.3in]{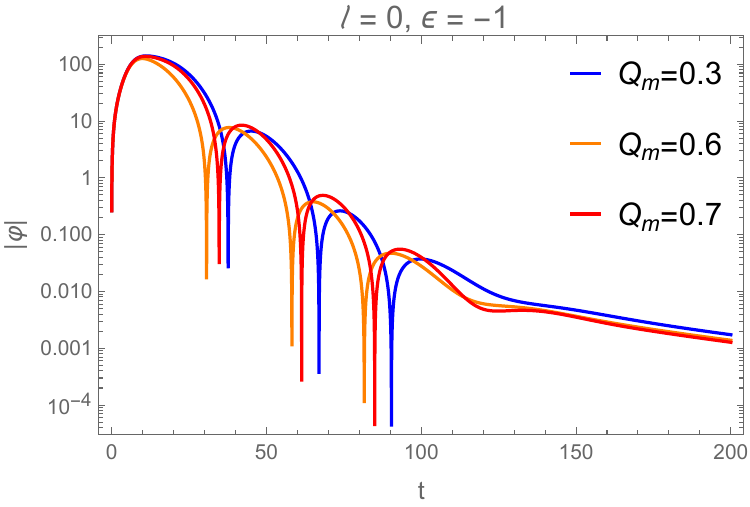}}
\subfigure[Perturbed electromagnetic field]{\label{figQ33} 
\includegraphics[width=2.3in]{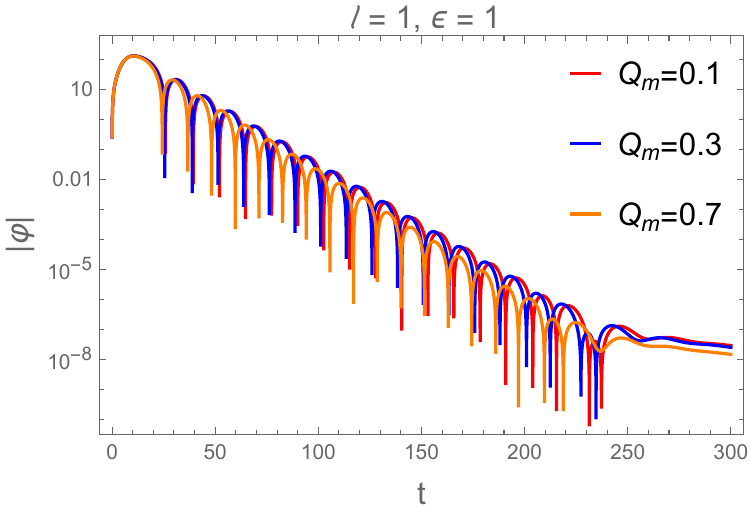}}
\subfigure[Perturbed electromagnetic field]{\label{figQ34} 
\includegraphics[width=2.3in]{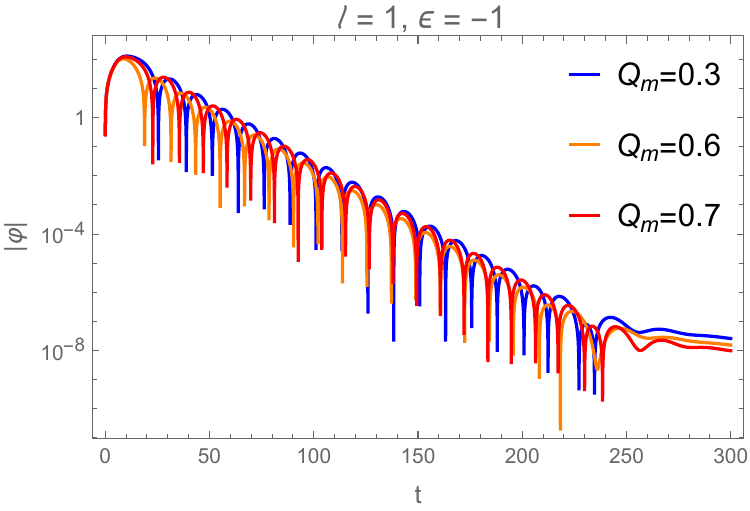}}
\caption{Time evolution for the perturbed scalar and electromagnetic fields with $M=1$. }\label{figQ3}
\end{figure}

In addition, we check how the magnetic charge $Q_m$ influences the evolution and waveform of various perturbations. 
For more details, the readers can refer to the Appendix C. The results for the lowest-lying modes are shown in Fig.\ref{figQ3}.
For the test scalar field perturbation on the black hole with $\epsilon=1$ (see Fig.\ref{figQ31}),
we observe that large $Q_m$ makes the ringing stage of perturbations waveform more intensive and shorter, which corresponds to the larger $Re(\omega)$ and  absolute value of $Im(\omega)$ (see Figs.\ref{figQ11}). 
In Figs. \ref{figQ32} and \ref{figQ34}, the perturbation wave with $Q_m=0.6$ perform a shorter ringing stage, indicating larger values of absolute value of $Im(\omega)$. This observation in time domains agree well with the results we obtained in the frequency domain for the black hole with $\epsilon=-1$, as shown in Figs. \ref{figQ14} and \ref{figQ24}.

\subsection{\textbf{\texorpdfstring{$\epsilon$}--dependence}}

\begin{figure}[H]
\centering
\subfigure[$\epsilon>0$]
{\label{figeps11} 
\includegraphics[width=2.3in]{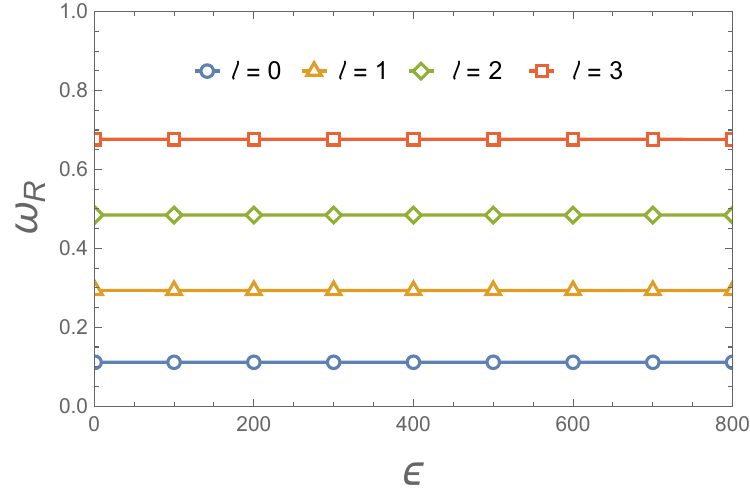}}
\subfigure[$\epsilon>0$]
{\label{figeps12} 
\includegraphics[width=2.4in]{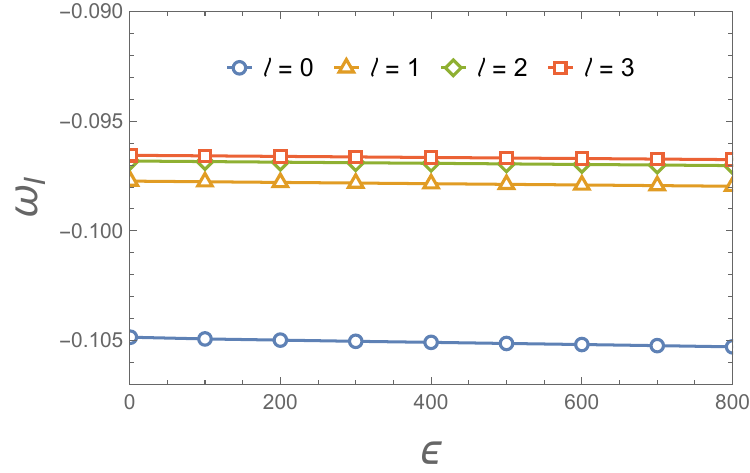}}
\subfigure[$\epsilon<0$]
{\label{figeps13} 
\includegraphics[width=2.3in]{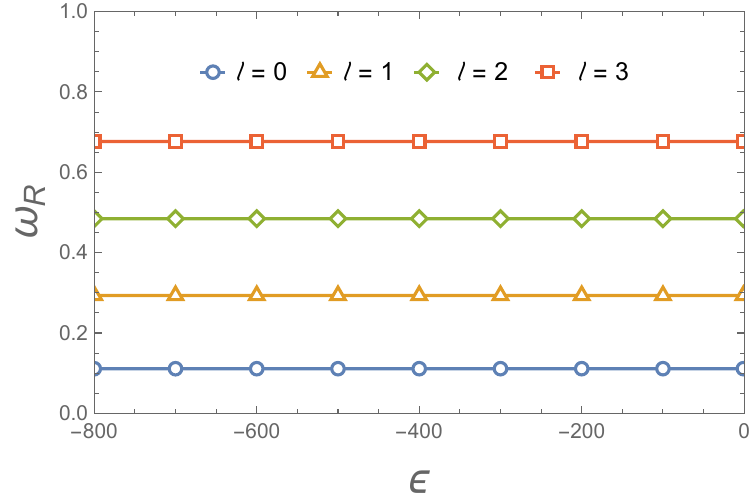}}
\subfigure[$\epsilon<0$]
{\label{figeps14} 
\includegraphics[width=2.4in]{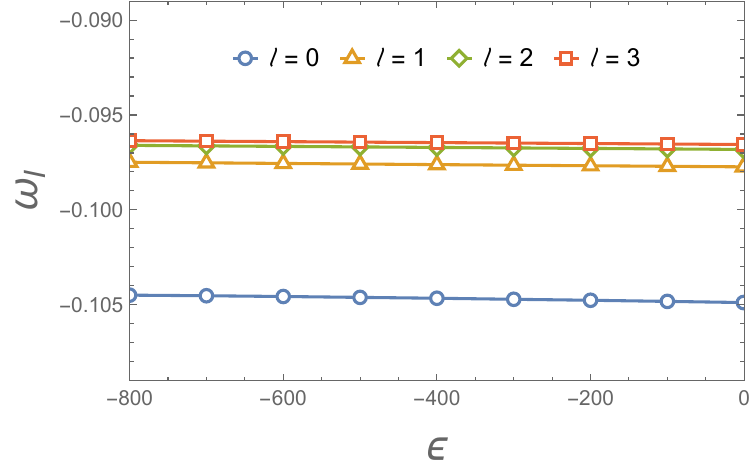}}
\caption{Variation of scalar fundamental QNFs with respect to the \textbf{different} $\epsilon$ with $Q_m=0.1$ and $M=1$.}\label{figeps1}
\end{figure}

\begin{figure}[H]
\centering
\subfigure[$\epsilon>0$]
{\label{figeps21} 
\includegraphics[width=2.3in]{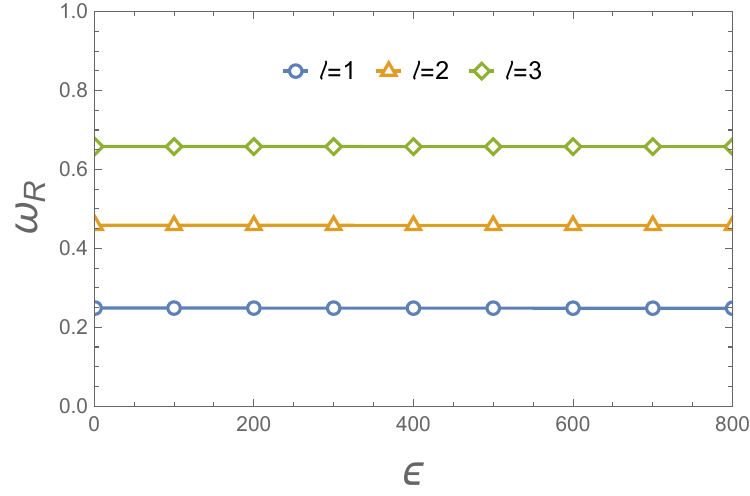}}
\subfigure[$\epsilon>0$]
{\label{figeps22} 
\includegraphics[width=2.4in]{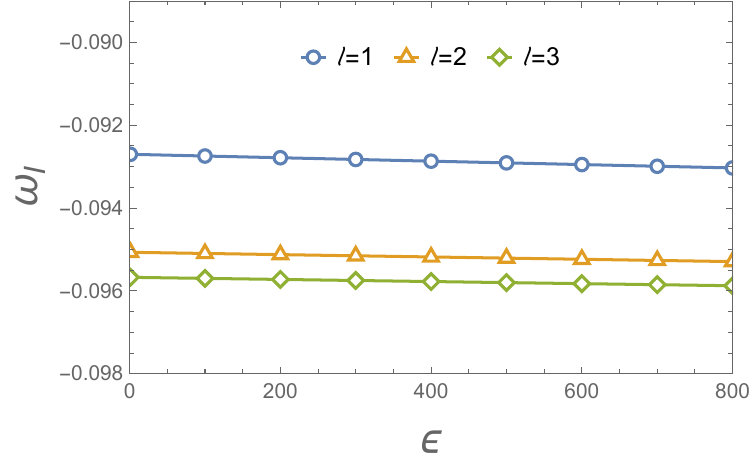}}
\subfigure[$\epsilon<0$]
{\label{figeps23} 
\includegraphics[width=2.3in]{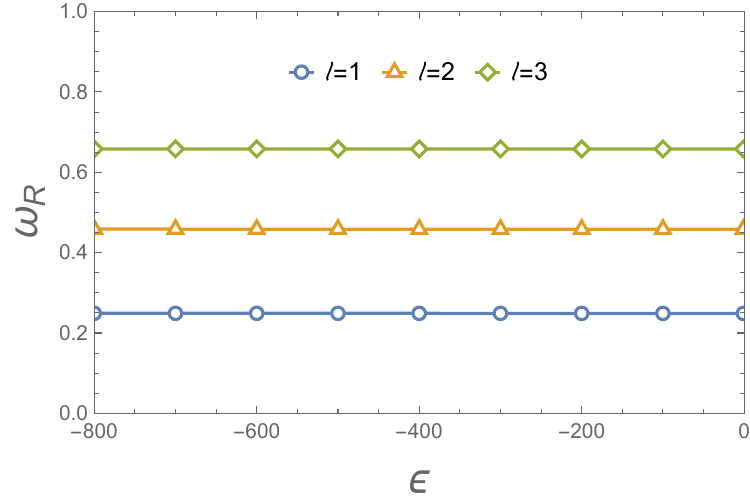}}
\subfigure[$\epsilon<0$]
{\label{figeps24} 
\includegraphics[width=2.4in]{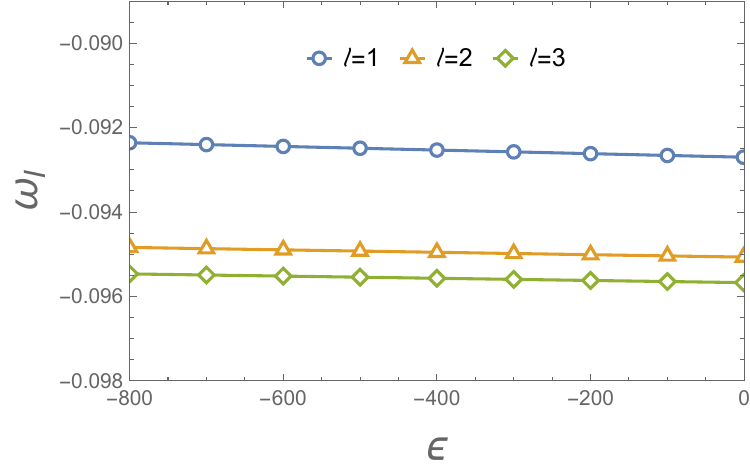}}
\caption{Variation of electromagnetic fundamental QNFs with respect to the \textbf{different} $\epsilon$ with  $M=1$.}\label{figeps2}
\end{figure}

\begin{figure}[H]
\centering
 \subfigure[Perturbed scalar field]{\label{fig:subfig:eps31} 
\includegraphics[width=2.3in]{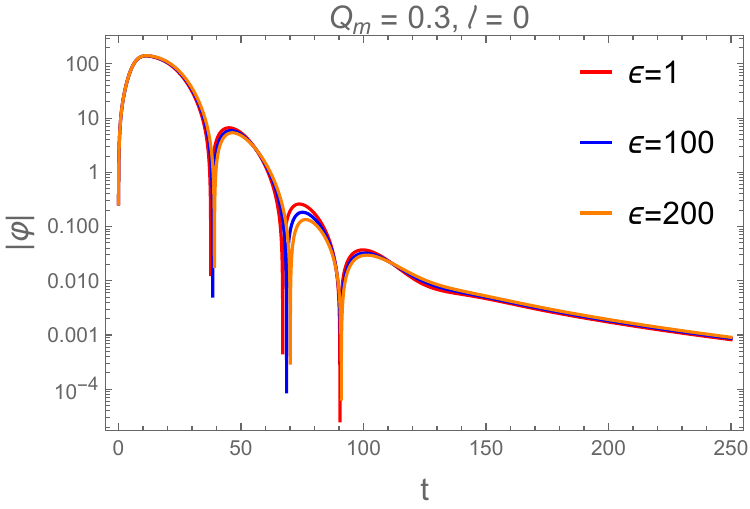}}
 \subfigure[Perturbed scalar field]{\label{fig:subfig:eps32} 
\includegraphics[width=2.3in]{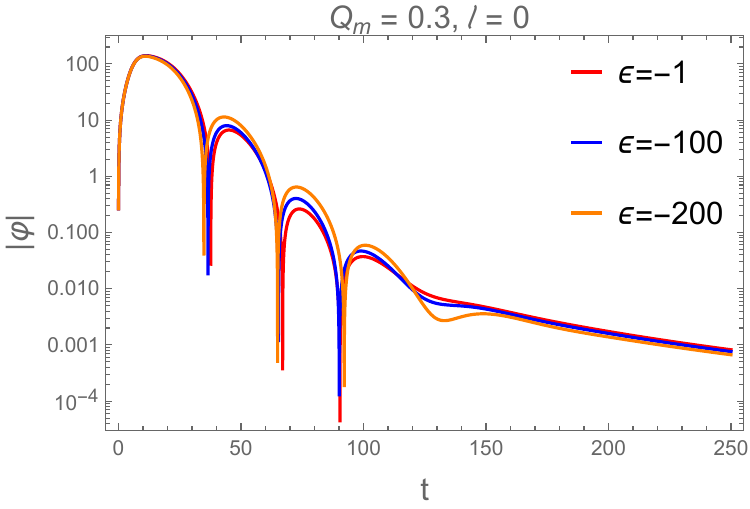}}
 \subfigure[Perturbed electromagnetic field]{\label{fig:subfig:eps33} 
\includegraphics[width=2.3in]{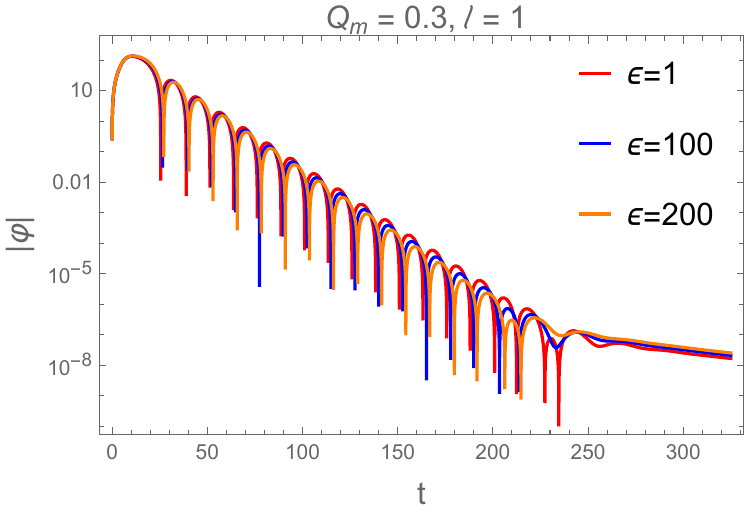}}
 \subfigure[Perturbed electromagnetic field]{\label{fig:subfig:eps34} 
\includegraphics[width=2.3in]{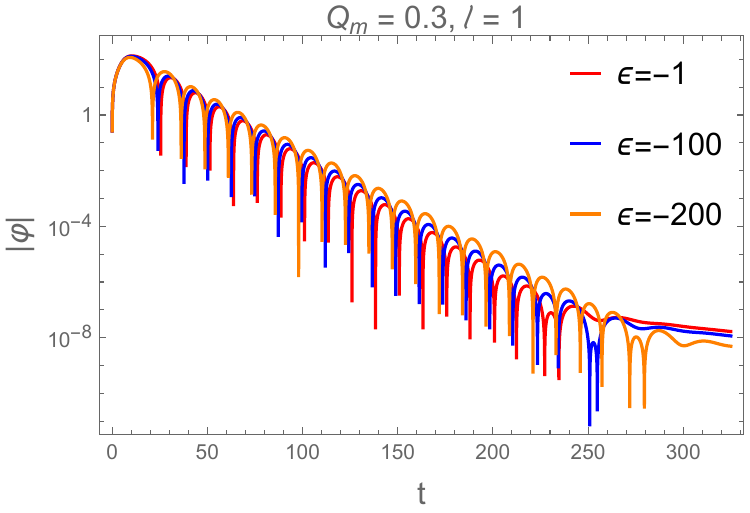}}
\hfill
\caption{Time evolution for perturbed scalar and electromagnetic fields with $M=1$. }\label{figeps3}
\end{figure}

It's also interesting to investigate the effect of  parameter $\epsilon$ on the QNFs for scalar and electromagnetic perturbations. Fixed the magnetic charge $Q_m=0.1$, we display the variation of fundamental QNFs and damping rate with different values of $\epsilon$ in Figs.\ref{figeps1} and \ref{figeps2}. As the parameter $\epsilon$ vary, the real and imaginary parts of these QNFs remain almost unchanged. These behaviors correspond to the potential function graphs (see Figs.\ref{fig13} and \ref{fig23}), including the propagation of these test fields in time domain (see Fig.\ref{figeps3}).

\section{ The Greybody factor}
\label{sec5}

In this section, we use the WKB approximation method to calculate greybody factors for scalar perturbation. 
The boundary condition for the scattering process is different from that of the QNMs, which can be written as
\begin{eqnarray}
\psi& =& T(\omega)e^{-i\omega r_*}, \quad r_* \rightarrow -\infty\nonumber,\\
\psi& =& e^{-i\omega r_*} + R(\omega)e^{i\omega r_*}, \quad r_* \rightarrow +\infty,\label{boundgrey}
\end{eqnarray}
where $R$ and $T$ represent the reflection coefficient and transmission coefficient, respectively. The greybody factor is defined as the probability of an outgoing wave
reaching to infinity or an incoming wave absorbed by the black hole. Therefore, $|T(\omega)|^2$ is called the greybody factor, and $R(\omega)$ and $T(\omega)$ should satisfy the following relation
\begin{eqnarray}
|R(\omega)|^2+|T(\omega)|^2=1.
\end{eqnarray}

Using the 6th-order WKB method, the
reflection and transmission coefficients can be obtained
\begin{eqnarray}
&&|R(\omega)|^2 = \frac{1} {1 + e^{-2\pi i K(\omega)}} ,\nonumber\\
&&|T(\omega)|^2 =\frac{1} {1 + e^{2\pi i K(\omega)}}= 1 -|R(\omega)|^2,\label{TR}
\end{eqnarray}
where $K$ is a parameter which can be obtained by the WKB formula
\begin{eqnarray}
K= \frac{i\left( \omega^2 - V(r_0) \right)}{\sqrt{-2V''(r_0)}} + \sum_{i=2}^6 \Lambda_i.
\end{eqnarray}

\begin{figure}[H]
\centering
\subfigure[$\epsilon=1$]{\label{fig1g}
\includegraphics[width=2.3in]{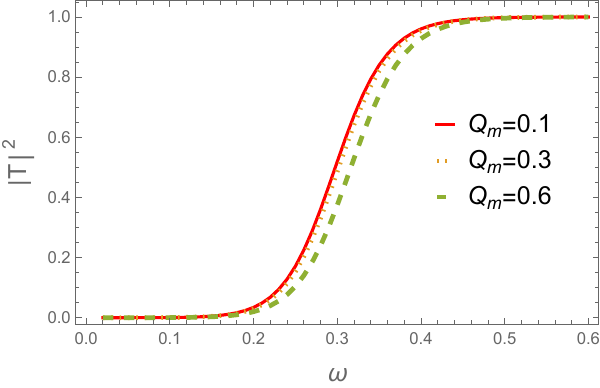}}
\subfigure[$\epsilon=-1$]{\label{fig2g} 
\includegraphics[width=2.3in]{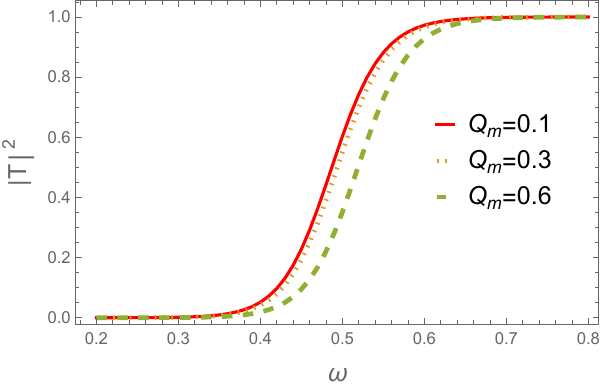}}
\subfigure[$Q_m=0.3$]{\label{fig3g}
\includegraphics[width=2.3in]{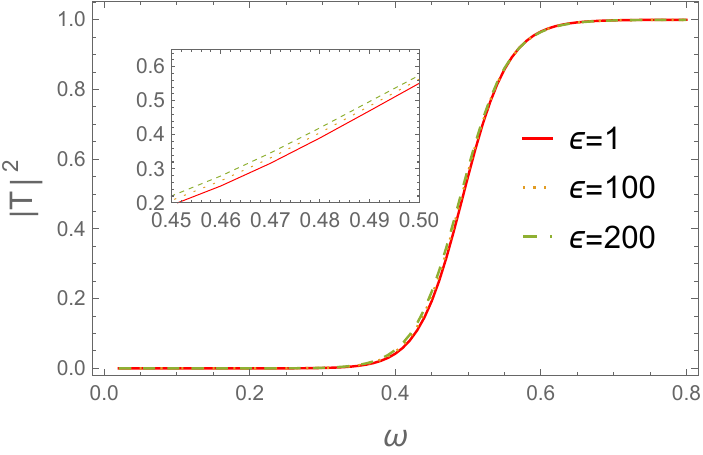}}
\subfigure[$Q_m=0.3$]{\label{fig4g} 
\includegraphics[width=2.3in]{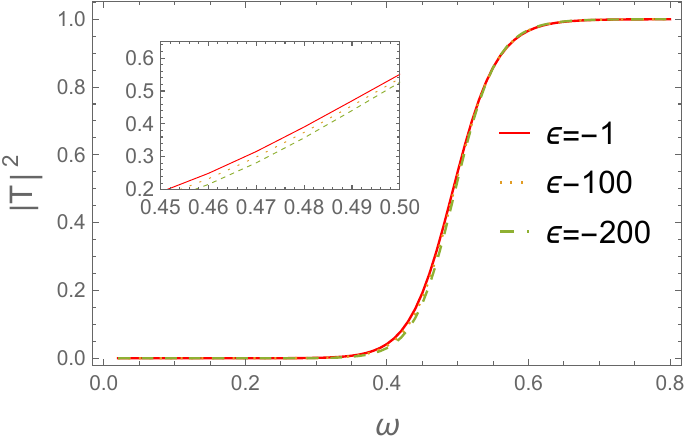}}
\caption{The scalar field greybody factor \textit{vs.} $Q_m$ with $M=1$.}\label{figs1}
\end{figure}

\begin{figure}[H]
\centering
\subfigure[$\epsilon=1$]{ \label{figs2a}
\includegraphics[width=2.3in]{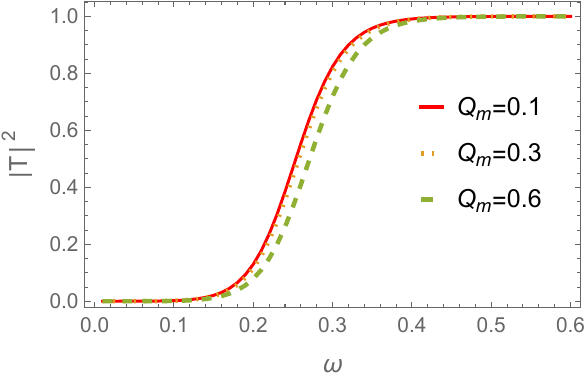}}
\subfigure[$\epsilon=-1$]{ \label{figs2b}
\includegraphics[width=2.3in]{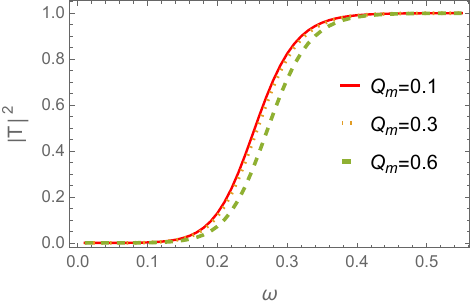}}
\subfigure[$Q_m=0.3$]{\label{figs2c} 
\includegraphics[width=2.3in]{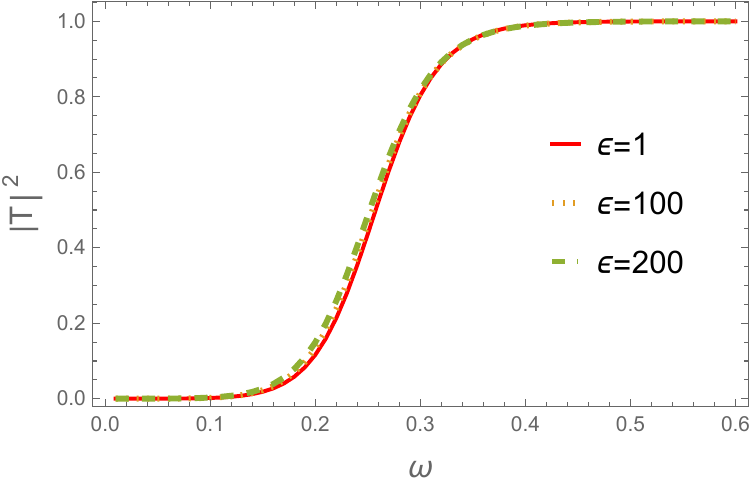}}
\subfigure[$Q_m=0.3$]{ \label{figs2d}
\includegraphics[width=2.3in]{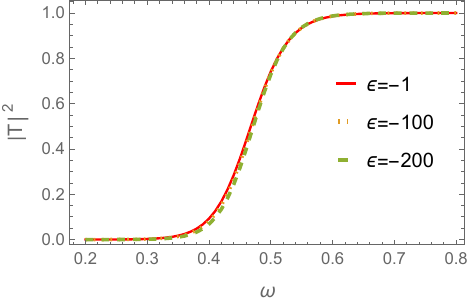}}
\caption{The electromagnetic field greybody factor \textit{vs.} $\epsilon$ with $M=1$.}\label{figs2}
\end{figure}

The figures presented in \ref{figs1} and \ref{figs2} provide a detailed analysis of the behavior of the greybody factors for massless
scalar and electromagnetic perturbations under varying the magnetic charge $Q_m$ and parameter $\epsilon$. One can see that the greybody factors exhibit noticeable decrease as $Q_m$ increases, which means that a smaller fraction of the perturbed field can penetrate the potential
barrier. We further present the influence of the difference $\epsilon$ on the greybody factors in Figs.\ref{fig3g}, \ref{fig4g}, \ref{figs2c} and \ref{figs2d}. As $\epsilon$ increases, the greybody factors also gradually increase.
It indicates that black holes become less interactive with the surrounding radiation and allow more of perturbed field to escape. These are all consistent with the effective potentials in Figs.\ref{fig1}- \ref{fig4}.

\section{Conclusion and discussion}
\label{sec6}

In this paper, we discussed the perturbations of different test fields on magnetically charged black holes in string-inspired Euler-Heisenberg theory, which involves a scalar field $\phi$ coupled to the electromagnetic field via a function $f(\phi)$. We found that this black hole solution reduces the form of GMGHS or GHS black holes, as $\epsilon$ tends toward 0. Moreover, this black hole possesses different horizon structures in cases of $\epsilon>0$ and $\epsilon<0$. The influences of magnetic charge $Q_m$, parameter $\epsilon$ and angular quantum number $l$ on the corresponding effective potentials of the perturbations have been analysed.

Then, we adopted the AIM and WKB methods to compute QNFs in each scenario.
To improve the numerical accuracy, the WKB approximation was extended to 6th order. We considered how QNFs changes with the magnetic charge $Q_m$ and difference $\epsilon$. For $\epsilon=1$, we found that the real QNFs increase as $Q_m$ increases, and the damping rate or decay rate of gravitational waves increases significantly with increase of $Q_m$. With regard to $\epsilon=-1$,
the real QNFs also increase as $Q_m$ increases. However, the damping rate increase with the growth of $Q_m$, and then reach a maximum decay rate occurring at $Q_m=0.6$. As $Q_m$ surpasses this point, the damping rate slowly begins to decrease. On the other hand,  we found the real part and damping rate of fundamental QNFs almost hold constant values when considering variation of difference $\epsilon$. These observations in frequency domains agree well with the results we obtained in the time domains for the two branch black hole solutions.


Using the 6th order WKB method, we also calculated the greybody factor for the perturbed scalar field. The results are shown in Figs.\ref{figs1} and \ref{figs2}. The effect of the magnetic charge $Q_m$ shows that
 a smaller fraction of the perturbed field can penetrate the potential
barrier. Conversely, an opposing phenomenon emerges under the variation of difference $\epsilon$. These are all consistent with the effective potentials.

 \vspace{1cm}
{\bf Acknowledgments}
 \vspace{0.5cm}

We gratefully acknowledge support by the National Natural Science Foundation of China (NNSFC) (Grant No.12365009), Jiangxi Provincial Natural Science Foundation (Grant No. 20232BAB201039) and Natural Science Basic Research Program of Shaanxi Province (Program No.2023-JC-QN-0053).

\appendix
\section{Asymptotic iteration method}

In this appendix, we will show the main steps of the AIM method. Firstly, we rewrite scalar perturbed equation \eqref{E} in terms of $u=1-r_+/r$
\begin{eqnarray}
  &&\psi''(u)+\frac{1}{2} \left(\frac{A'(u)}{A(u)}+\frac{B'(u)}{B(u)}+\frac{4}{u-1}\right) \psi'(u) \nonumber\\
  &&+\left[\frac{r_+^2 \omega ^2}{(u-1)^4 A(u) B(u)}+\frac{A'(u)}{2 (u-1) A(u)}+\frac{B'(u)}{2 (u-1) B(u)}-\frac{l (l+1)}{(u-1)^2 B(u)} \right]\psi(u)=0.\label{equ1}
\end{eqnarray}
Evidently, the range of $u$ satisfies $0\leqslant u<1$. 

From \eq{equ1}, we consider the behavior of the function $\psi(u)$ at horizon $(u=0)$ and at the boundary $u=1$. Near the horizon $(u=0)$, we have $A(0)\approx u A'(0)$ and $B(0)\approx u B'(0)$. Then \eq{equ1} becomes
\begin{eqnarray}
  \psi''(u)+\frac{1}{u}\psi'(u)+\frac{r_+^2\omega^2}{u^2 A'(0) B'(0)}\psi(u)=0.
\end{eqnarray}
We can obtain the solution
\begin{eqnarray}
  \psi(u\to 0)\sim C_1 u^{-\xi}+C_2 u^{\xi},
\end{eqnarray}
with $\xi=\frac{ir_+\omega}{\sqrt{A'(0) B'(0)}}$.
Then we have to set $C_2=0$ in order to respect the ingoing condition at the black hole horizon.

At infinity $(u=1)$, the asymptotic form of \eq{equ1} can be written as
\begin{eqnarray}
  \psi''(u)-\frac{2}{1-u}\psi'(u)+\frac{r_+^2\omega^2}{(1-u)^4}\psi(u)=0,
\end{eqnarray}
with $A(1)=1$ and $B(1)=1$.
Then, we can obtain the solution
\begin{eqnarray}
  \psi(u\to 1)\sim D_1 e^{-\zeta}+D_2 e^{\zeta},~ \zeta=\frac{ir_+\omega}{1-u}.
\end{eqnarray}
In order to impose the outgoing boundary condition, we should set $D_1=0$.

Now, using the above solutions at horizon and infinity, we can define the general ansatz for \eq{equ1} as
\begin{eqnarray}
  \psi(u)=u^{-\xi}e^{\zeta}\chi(u) \label{eqschi}.
\end{eqnarray}
Substitute \eq{eqschi} to \eq{equ1}, we have
\begin{eqnarray}
  \chi''=\lambda_0(u)\chi'+s_0(u)\chi \label{eqschieq},
\end{eqnarray}
where 
\begin{eqnarray}
  \lambda_0(u)=\frac{1}{2} \left(\frac{4 i r_+ \omega }{u \sqrt{A'(0)} \sqrt{B'(0)}}-\frac{A'(u)}{A(u)}-\frac{B'(u)}{B(u)}-\frac{4 \left(i r_+ \omega +u-1\right)}{(u-1)^2}\right),
\end{eqnarray}
and
\begin{eqnarray}
  s_0(u)&=&\frac{1}{2 (u-1)^4 u^2 A(u) A'(0) B(u) B'(0)}\Big[ i r_+ (u-1)^4 u \omega  \sqrt{A'(0)} B(u) \sqrt{B'(0)} A'(u) \nonumber\\
  &&-u^2 A'(0) B'(0) \left(2 r_+^2 \omega ^2+(u-1)^2 B(u) A'(u) \left(i r_+ \omega +u-1\right)\right)  \nonumber\\
  &&+A(u) \Bigg(2 l (l+1) u^2 (u-1)^2 A'(0) B'(0)
  -u^2 (u-1)^2 A'(0) B'(0) B'(u) \left(i r_+ \omega +u-1\right)\nonumber\\
  &&+i r_+ u (u-1)^4 \omega  \sqrt{A'(0)} \sqrt{B'(0)} B'(u)+2 r_+ \omega  B(u) \Big(r_+ \omega  \left((u-1)^2-u \sqrt{A'(0)} \sqrt{B'(0)}\right)^2\nonumber\\
  &&+i (u+1) (u-1)^3 \sqrt{A'(0)} \sqrt{B'(0)}\Big)\Bigg) \Big].
\end{eqnarray}
Based on $\lambda_0$ and $s_0$, the perturbed equation \eqref{eqschieq} can be solved numerically by using the improved AIM \cite{Cho:2009cj}. 

Following the same procedure described above, we can also obtain functions  $\lambda_0$ and  $s_0$  for the electromagnetic field perturbation:  
\begin{eqnarray}
  \lambda_0(u)=\frac{1}{2} \left(\frac{4 i r_+ \omega }{u \sqrt{A'(0)} \sqrt{B'(0)}}-\frac{A'(u)}{A(u)}-\frac{B'(u)}{B(u)}-\frac{4 \left(i r_+ \omega +u-1\right)}{(u-1)^2}\right),
\end{eqnarray}
and
\begin{eqnarray}
  s_0(u)&=&\frac{1}{2 (u-1)^4 u^2 A(u) A'(0) B(u) B'(0)}\Big[r_+ (u-1)^4 u \omega B(u) \sqrt{A'(0)} A'(u) \sqrt{B'(0)} \nonumber\\
&&- r_+ u^2 \omega A'(0) \left(2 r_+ \omega + i ( u-1)^2 B(u) A'(u)\right) B'(0)  \nonumber\\
&&+ A(u) \Bigg(2 r_+ \omega B(u) \bigg(r_+ \omega \left((  u-1)^2 - u \sqrt{A'(0)} \sqrt{B'(0)}\right)^2 \nonumber\\
&& + i ( u-1)^3 (1 + u) \sqrt{A'(0)} \sqrt{B'(0)}\bigg)+2 l (1 + l) ( u-1)^2 u^2 A'(0) B'(0) \nonumber\\
&& + i r_+ (u-1)^2 u \omega \sqrt{A'(0)} \left((u-1)^2 - u \sqrt{A'(0)} \sqrt{B'(0)}\right) \sqrt{B'(0)} B'(u)\Bigg)\Big].
\end{eqnarray}

\section{6th-order WKB method}

This method, first proposed by Schutz and Will, was used to address black hole scattering problems \cite{Kokkotas:1988fm}. Later, further developments were made by Iyer, Will, and Konoplya  \cite{Konoplya:2011qq}. In this paper, we consider the most commonly used 6th-order WKB approximation method \cite{Konoplya:2011qq}
\begin{align}
\frac{i(\omega^2 - V_0)}{\sqrt{-2V_0''}} - \sum_{i=2}^6 \Lambda_i = n + \frac{1}{2}, \quad (n = 0, 1, 2, \cdots)
\end{align}
where \( V''(r_0) \) is the value of the second derivative of the effective potential with respect to \( r \) at its maximum point \( r_0 \) defined by the solution of the equation \( \left. \frac{dV}{dr_*} \right|_{r_*=r_0}= 0 \). \( V_0 \) represents the maximum value of the effective potential, and \( \Lambda_i \) is the \( i \)-th order revision terms depending on the values of the effective potential. This semi-analytical method has been applied extensively in numerous black hole spacetime cases. 
It should be pointed out here the WKB approach works well for situations where the multipole number is larger compared to the overtone: $l\geq n$, the WKB approach produces less accurate outcomes for $l<n$ \cite{Iyer:1986nq,Konoplya:2003ii}.

\section{Time domain integration}

In order to illustrate the properties of QNMs from the propagations of various fields, we shall shift our analysis into the time domain. To this end, we reconstruct the Schrodinger-like equations (\eqref{E},\eqref{e}) into the time-dependent form by simply replacing the second-order term with $-\frac{d^2}{dt^2}$, then we have the uniformed second-order partial differential equation for various perturbed fields as
\begin{eqnarray}
\left ( \frac{d^2}{dr_*}-\frac{d^2}{dt^2} -V(r)\right )\Psi(r,t)=0.
\end{eqnarray}

To solve the above equations, one has to deal with the time-dependent evolution problem. A convenient way is to adopt the finite difference method\cite{Abdalla:2010nq} to numerically integrate these wave-like equations at the time coordinate and fix the space configuration with a Gaussian wave as an initial value of time. To handle this, one firstly discretizes the radial coordinate with the use of the definition of tortoise coordinate
\begin{eqnarray}
\frac{dr (r_{*})}{dr_{*}} &=& \sqrt{A(r(r_*))B(r(r_*))}\Rightarrow \frac{r(r_{*j} + \Delta r_{*}) - r (r_{*j})}{\Delta r_{*}}\nonumber\\
& = &\frac{r_{j+1}-r_{j}}{\Delta r_{*}} = \sqrt{A(r_j)B(r_j)} \Rightarrow r_{j+1} = r_{j} + \Delta r_{*}\sqrt{A(r_j)B(r_j)} .
\end{eqnarray}

Then one can further discretize the effective potential into
$V(r(r_*)) = V(j\Delta r_*) = V_j$ and the field into $\Psi(r, t) =
\Psi(j\Delta r_*, i\Delta t) = \Psi_{j,i}$. Subsequently, the wave-like equation \cite{Zerilli:1970se} turns out to be a discretized equation
\begin{eqnarray}
-\frac{\Psi_{j,i+1} - 2\Psi_{j,i} + \Psi_{j,i-1}}{\Delta t^2} + \frac{\Psi_{j+1,i} - 2\Psi_{j,i} + \Psi_{j-1,i}}{\Delta r_*^2}
-V_j\Psi_{j,i} + \mathcal{O}(\Delta t^2) + \mathcal{O}(\Delta r_*^2) = 0,
\end{eqnarray}
from which one can isolate $\Psi_{j,i+1}$ after algebraic operations
\begin{eqnarray}
\Psi_{j,i+1} = \frac{\Delta t^2}{\Delta r_*^2}\Psi_{j+1,i} + \left(2 - 2\frac{\Delta t^2}{\Delta r_*^2} - \Delta t^2 V_j\right) \Psi_{j,i} + \Psi_{j-1,i} - \Psi_{j,i-1}.
\end{eqnarray}

The above equation is nothing but an iterative equation, which can be solved if one gives a Gaussian wave packet $\Psi_{j,0}$ as the initial perturbation. In our calculations,  setting the seed $r_{i=1}=r_h+10^{-12}$, after imposing the initial condition $\Psi_{j,i<0} = 0$, $\Psi_{j,0}=\exp[-\frac{(r_{j}-a)^2}{2b^2}]$. We choose the parameters $a = 0.1$ and $ b = 3$ in the Gaussian profile and setting $\frac{\Delta t}{\Delta r_*}=\frac{0.025}{0.05}=\frac{1}{2}$, this iterative equation provides us the evolution of the scalar field and electromagnetic field perturbation in time profile.

\end{document}